\documentclass[prb,reprint,twocolumn,amsmath,amssymb]{revtex4-1}

\newcommand{\br}{\boldsymbol{\textbf{r}}}

\newcommand{\bx}{\boldsymbol{\textbf{x}}}

\newcommand{\bR}{\boldsymbol{\textbf{R}}}

\newcommand{\dx}{\,d\bx}

\newcommand{\order}{\mathcal{O}}

\usepackage{graphicx}
\usepackage{graphics}
\usepackage{multirow}
\usepackage{fancyhdr, ifpdf}
\usepackage{amsxtra}
\usepackage{stmaryrd}
\usepackage{mathrsfs}
\usepackage{psfrag}
\usepackage{subfig}
\usepackage{epsfig}
\usepackage{rotating}
\usepackage{setspace}
\usepackage{float}
\usepackage{amsfonts}
\usepackage{amsbsy}
\usepackage{amscd}
\usepackage{amsthm}
\usepackage{color}
\usepackage{tabularx}
\usepackage{caption}
\usepackage{sidecap}
\usepackage{dcolumn}
\usepackage{natbib}
\usepackage{footnote}
\usepackage{algorithm}
\usepackage{algorithmic}
\usepackage[normalem]{ulem}
\usepackage[titletoc,title]{appendix}
\usepackage{hyperref}

\definecolor{hellgruen}{rgb}{0.2,0.7,0.2}

\newcolumntype{M}[1]{>{\centering\arraybackslash}m{#1}}
\newcolumntype{N}{@{}m{0pt}@{}}
\begin{document}
\preprint{}
\title{Large-scale all-electron density functional theory calculations using an enriched finite element basis}

\author{Bikash Kanungo}
\author{Vikram Gavini}

\affiliation{Department of Mechanical Engineering, University of Michigan, Ann Arbor, MI 48109, USA}

\begin{abstract}
We present a computationally efficient approach to perform large-scale all-electron density functional theory calculations by enriching the classical finite element basis with compactly supported atom-centered numerical basis functions that are constructed from the solution of the Kohn-Sham (KS) problem for single atoms. We term these numerical basis functions as enrichment functions, and the resultant basis as the enriched finite element basis. The compact support for the enrichment functions is obtained by using smooth cutoff functions, which enhances the conditioning and maintains the locality of the enriched finite element basis. The integrals involved in the evaluation of the discrete KS Hamiltonian and overlap matrix in the enriched finite element basis are computed using an adaptive quadrature grid that is constructed based on the characteristics of enrichment functions. Further, we propose an efficient scheme to invert the overlap matrix by using a block-wise matrix inversion in conjunction with special reduced-order quadrature rules, which is required to transform the discrete Kohn-Sham problem to a standard eigenvalue problem. Finally, we solve the resulting standard eigenvalue problem, in each self-consistent field iteration, by using a Chebyshev polynomial based filtering technique to compute the relevant eigenspectrum. We demonstrate the accuracy, efficiency and parallel scalability of the proposed method on semiconducting and heavy-metallic systems of various sizes, with the largest system containing 8694 electrons. We obtain accuracies in the ground-state energies that are $\sim1\,$mHa with reference ground-state energies employing classical finite element as well as gaussian basis sets. Using the proposed formulation based on enriched finite element basis, for accuracies commensurate with chemical accuracy, we observe a staggering $50-300$ fold reduction in the overall computational time when compared to classical finite element basis. Further, we find a significant outperformance by the enriched finite element basis when compared to the gaussian basis for the modest system sizes where we obtained convergence with gaussian basis. We also observe good parallel scalability of the numerical implementation up to 384 processors for a representative benchmark system comprising of 280-atom silicon nano-cluster. 
\end{abstract}

\maketitle

\section{Introduction}\label{sec:Intro}

Kohn-Sham Density Functional Theory (DFT), enjoying the distinction of the most widely used electronic structure method for over four decades, has immensely contributed to our understanding of a wide range of materials properties. It relies on the Hohenberg-Kohn theorem ~\cite{Hohenberg1964} and the Kohn-Sham \textit{anstaz} ~\cite{Kohn65} to reduce the many-body Schr\"odinger equation to an effective single electron problem, thereby, making predictions on materials properties computationally tractable. On the other hand, the pseudopotential approximation~\cite{Payne1992,Bachelet1982,Chelikowsky2000,Schwerdtfeger2011b} has played an important role in electronic structure method development, which reduces the electronic structure calculation to the evaluation of smooth pseudo-wavefunctions corresponding to the valence electrons of a Hamiltonian constructed from a smooth effective external potential, namely the pseudopotential. The construction of a pseudopotential, which is non-unique, entails matching the pseudo-wavefunctions to the corresponding all-electron wavefunctions outside the user defined atomic core. In the past few decades, pseudopotentials have seen a rapid evolution from norm-conserving potentials~\cite{Hamann1979, Vanderbilt1985, Hamann1989, Troullier1991} to ultrasoft potentials ~\cite{Vanderbilt1990} to the state-of-the-art projector augmented wave (PAW) ~\cite{Blochl1994} method and have proven to be successful in predicting bulk, mechanical, electrical, magnetic, and chemical properties for a wide range of materials.

However, despite their success, pseudopotentials  are often sensitive to the choice of core size used in their construction and tend to oversimplify the treatment of core electrons as chemically inert for various systems and external conditions. For example, in systems under high pressure where the core and valence wavefunctions show increasing overlap with pressure, pseudopotentials tend to underpredict their phase transition pressures ~\cite{Oganov2003,Abu-Jafar2000,Xiao2010}; in systems at high temperature, where the contribution of core electrons to various thermodynamic potentials is non-negligible, pseudopotentials provide an inaccurate description of the equation of state ~\cite{Levashov2010}; in transition metals, where the penultimate $d$ and $f$ orbitals are not tightly bound, non-inclusion of these orbitals as valence electrons oftentimes lead to inaccurate bulk property prediction. More pronounced inaccuracies and sensitivity to core sizes are observed in prediction of ionization potentials ~\cite{Liu1998}, magnetizatibility ~\cite{Schwerdtfeger2011a}, spectroscopic properties ~\cite{Schwerdtfeger1995, Schwerdtfeger2000} of heavier atoms wherein scalar relativistic pseudopotentials are widely employed, and in prediction of band-gap and excited state properties ~\cite{Gomez-Abal2008}. Thus, all-electron calculations are necessary for an accurate and more reliable description of such systems and conditions.

The earliest and the most commonly employed method for all-electron calculations involves the use of atomic-orbital-type basis sets~\cite{Hehre1969, Gaussian2009, TurboMol1989, Gamess1993, QChem2006, FhiAims2009, Nwchem2010, MolPro2012, Dalton2014}, wherein atom-specific basis, either analytic or numerical, are used with only a few basis functions per atom. However, owing to the incompleteness of the basis, systematic convergence for all materials systems remains a concern. Moreover, in many numerical implementations, their applicability is largely limited to isolated systems and are not easily amenable to arbitrary boundary conditions. Furthermore, the non-locality of the basis substantially limits parallel scalability of their numerical implementations. Among the family of complete basis sets, the plane-wave basis, owing to the straightforward evaluation of the Coulomb interactions in Fourier space and the exponential convergence afforded by the basis, has been the most popular choice for pseudopotential calculations. However, its applicability to all-electron calculations is greatly hindered by its lack of adaptive spatial resolution, as any computationally efficient basis for all-electron calculations warrants finer resolution closer to nuclei, where the wavefunctions are rapidly oscillating, and coarser resolution elsewhere. This shortcoming has been, to a large extent, alleviated through the use of various augmentation schemes such as Augmented Plane-wave (APW) ~\cite{Loucks1967, Koelling1975}, Linearized Augmented Plane-wave (LAPW) ~\cite{Andersen1975, Wimmer1981, Weinert1982} and APW+lo ~\cite{Sjostedt2000, Madsen2001, Gulans2014}. All these methods involve the description of the wavefunctions in terms of products of radial functions and spherical harmonics inside muffin-tins (MTs) surrounding each atom, and in terms of plane-waves in the interstitial regions between atoms. Although these schemes attain adaptive spatial resolution, the basis functions within the MTs depend on the choice of trial energy parameters, typically based on atomic energies, for each azimuthal ($l$) quantum number. Owing to the lack of one-to-one correspondence between the Kohn-Sham eigenvalues and the trial energy parameters, the quality of the basis is sensitive to the choice trial energy parameters, especially in systems where the chosen $l$ quantum number based trial energies fail to describe all states with the same $l$-character, and in systems where the occupied bands show substantial deviation from their atomic counterparts ~\cite{Gulans2014}.  Additionally, certain notable disadvantages of plane-waves such as their restriction to periodic boundary conditions, the highly non-local communication associated with Fast Fourier Transform (FFT), also persist in these augmentation schemes.      

Bl\"ochl, in his PAW formulation ~\cite{Blochl1994}, generalized the notion of APW/LAPW and the pseudopotential approach to construct the all-electron orbitals through a linear transformation, $\hat{\mathscr{T}}$, of the smoothly varying pseudo orbitals, thus providing a balance between accuracy and computational efficiency. However, typically, PAW is implemented within the frozen-core approximation, wherein, although the oscillatory behavior of the valence orbitals near the atomic centers is retrieved through $\hat{\mathscr{T}}$ acting on the pseudo valence orbitals, the core states are treated as frozen and do not feature within the self-consistent field iteration. One can, in principle, relax the core states within the PAW framework, however, this involves achieving simultaneous self-consistency in core states, valence partial waves and the effective potential, which can severely affect the computational efficiency otherwise afforded by frozen-core approximation. Marsman et. al ~\cite{Marsman2006} presented a computationally efficient extension of PAW beyond the frozen-core approximation, wherein, first, the core states are updated self-consistently within a fixed valence charge density and a spherical approximation for the one-center potential. Subsequently, new valence partial waves are evaluated. However, as noted in that work, the spherical approximation of the one-center potential used in the core-state relaxation poses limitations in terms of accounting for core polarization effects and core-core interactions from neighboring atoms; capturing changes in valence-core interactions outside the augmentation spheres; preserving orthogonality of the valence partial waves with the core states under situations where the core charge density spills outside the augmentation spheres. Additionally, the construction of the valence all-electron and pseudo partial waves that feed into $\hat{\mathscr{T}}$, while using the actual one-center potential (crystal potential) in their construction, involves the use of trial energy parameters (analogous to APW/LAPW), thereby introducing sensitivity to the choice of these trial energies. Therefore, to account for these notable limitations, it is desirable to treat the core electrons on equal footing with the valence electrons while at the same time minimize the huge computational expense incurred by such explicit treatment of core electrons.

The limitations of plane-waves have, in the past two decades, led to the development of various real-space techniques for DFT calculations, of which the Finite Difference (FD) method ~\cite{Chelikowsky1994, Kronik2006} remains the most prominent. The FD method can handle arbitrary boundary conditions, and exhibit improved parallel scalability in comparison to plane-wave basis. However, the FD method fails to retain the variational convergence of plane-waves. Moreover, a lack of basis in the FD method makes an accurate treatment of singular potentials difficult, thereby, limiting its utility for all-electron calculations. Finite element basis ~\cite{Brenner2007, Hughes2012}, on the other hand, being a local piecewise polynomial basis, retains the variational property of the plane-waves, and, in addition, has other desirable features such as locality of the basis that affords good parallel scalability, being easily amenable to adaptive spatial resolution, and the ease of handling arbitrary boundary conditions. While most studies employing the finite element basis in DFT calculations~\cite{White1989, Tsuchida1998, Pask1999, Pask2001, Pask2005, Zhang2008, Suryanarayana2010, Fang2012, Bao2012, Motamarri2013} have shown its usefulness in pseudopotential calculations, some of the works~\cite{White1989, Bylaska2009, Lehtovaara2009, Motamarri2013, Schauer2013, Motamarri2014} have also demonstrated its promise for all-electron calculations. In particular, the work of Motamarri et. al ~\cite{Motamarri2013} has combined the use of higher-order spectral finite elements along with Chebyshev polynomial based filtering technique to develop an efficient scheme for the computation of the occupied eigenstates. As detailed in the work, the aforementioned method outperforms the plane-wave basis in pseudopotential calculations for the benchmark systems considered. However, in the context of all-electron calculations, it remains an order of magnitude slower in comparison to the gaussian basis. This relatively unsatisfactory performance of the finite element basis in all-electron calculations was attributed to the requirement of large number of basis functions ($\order(10^5)$ per atom, even for light atoms) as well as the high polynomial degree required in the Chebyshev filter ($\order(10^3)$) to accurately compute the occupied eigenstates. To elaborate, one requires a highly refined finite element mesh closer to the atomic cores in order to capture the sharp variations in the electronic wavefunctions and the singularity of the nuclear potential. This refinement, in turn, leads to an increase in the spectral width of the discrete Kohn-Sham Hamiltonian, thereby, requiring a very high polynomial degree Chebsyhev filter to compute the occupied eigenstates. This need for a high polynomial degree Chebyshev filter in all-electron calculations also negatively effects the computational complexity realized through reduced order scaling methods. As detailed in a recent work 
~\cite{Motamarri2014}, which combines Chebyshev filtered subspace projection with localization and Fermi-operator expansion, while pseudopotential calculations exhibited linear scaling for materials systems with a band-gap and subquadratic scaling for materials systems without a band gap, the overall scaling for all-electron calculations was close to quadratic even for materials with a band-gap. 


In order to alleviate the aforementioned limitations of finite element basis in all-electron calculations, we propose employing a mixed basis comprising of finite element basis functions and compactly supported atomic-orbital-type basis functions. In particular, the atomic-orbital-type functions capture the essential features of the electronic fields near the nuclei, thereby, mitigating the need for high mesh refinement around atoms, while the finite element basis functions capture the smooth parts of the wavefunction away from the nuclei and also extend completeness to the basis. In this work, we formalize this idea of a mixed basis to develop, what we refer to as, the enriched finite element basis. The enriched finite element basis is generated by augmenting the piecewise continuous Lagrange polynomials in finite element basis, henceforth described as the classical finite element basis, with compactly supported atom-centered numerical basis functions that are obtained from the solutions of the Kohn-Sham problem (Kohn-Sham orbitals and electrostatic potentials) for single atoms. We term these atom-centered numerical basis functions as enrichment functions. We note that the proposed enriched finite element basis differs from other augmentation schemes in plane-waves like APW, LAPW, and APW+lo in the following ways: (i) unlike the plane-wave augmentation schemes, the enriched finite element basis does not partition the space into muffin tins (MTs) and interstitials, thereby eliminating the need of any matching or smoothness constraint for the augmenting basis functions; (ii) as opposed to the plane-wave augmentation schemes, the enrichment functions of our proposed method do not have any trial energy parameter dependence; and (iii) unlike the plane-wave augmentation scheme, wherein the basis functions inside the MTs needs to be computed for every materials system separately, the enrichment functions, being atomic solutions to the electronic fields, are independent of the materials system and are computed \textit{a priori}.

The key ideas in the proposed method involve: (i) pre-computing the enrichment functions by solving radial Kohn-Sham equations and employing smooth cutoff functions to ensure the locality as well as control the conditioning of the enriched finite element basis; (ii) employing a divide and conquer strategy to construct an adaptive quadrature grid based on the nature of enrichment functions, so as to accurately and efficiently evaluate the integrals involving enrichment functions; (iii) implementing an efficient scheme to evaluate the inverse of the overlap matrix corresponding to the enriched finite element basis by using block-wise matrix inversion in conjunction with Gauss-Lobatto-Legendre reduced order quadrature rules; and (iv) in each self-consistent field iteration, using a Chebyshev polynomial based filter to compute the space spanned by the occupied eigenstates, and solving the Kohn-Sham eigenvalue problem by projecting the problem onto this Chebyshev-filtered space. We have implemented the proposed method in a parallel computing framework using the Message Passing Interface (MPI) to enable large-scale all-electron calculations. To begin with, we demonstrate optimal convergence rates of the ground-state energies with respect to enriched finite element basis. Further, we investigate the accuracy and performance of the proposed method on benchmark semi-conducting (silicon nano-clusters) and heavy-metallic (gold nano-clusters) systems of various sizes, with the largest system containing 8694 electrons. The proposed formulation using the enriched finite element basis obtains close to $1\,$mHa accuracy in per-atom ground-state energies of the benchmark systems when compared to the reference ground-state energies obtained from classical finite element basis or gaussian basis calculations. Furthermore, the proposed method achieves a staggering $50-300$ fold speedup relative to the classical finite element basis, and a significant speedup relative to the gaussian basis even for modest sized systems. Lastly, we observe good parallel efficiency of our implementation up to 384 processors for a silicon nano-cluster containing 3920 electrons discretized using $\sim 4$ million basis functions. 

The rest of the paper is structured as follows. In Section ~\ref{sec:KSVP} we recall the real-space formulation of the Kohn-Sham DFT problem. Subsequently, we briefly introduce the classical finite element discretization in the context of Kohn-Sham DFT problem in Section ~\ref{sec:CFEM}. Section~\ref{sec:EFEM} details the proposed enriched finite element discretization for the Kohn-Sham eigenvalue problem, and section ~\ref{sec:ChFSI} discusses the key ideas based on Chebyshev polynomial filtering employed in the self-consistent field iteration (SCF) solution procedure. Section ~\ref{sec:Results} presents the convergence, accuracy, performance and parallel scalability of the enriched finite element basis. Finally, we summarize the findings from the present work and outline the future scope in Section~\ref{sec:Summary}.

\section{Real-space DFT formulation}\label{sec:KSVP}
We recall that the ground-state properties of a materials system in the Kohn-Sham DFT framework are computed by solving the non-linear Kohn-Sham eigenvalue problem~\cite{Kohn65}, given by
\begin{equation} \label{eq:KSEig}
  \left(-\frac{1}{2}\nabla^{2}+V_{\text{eff}}(\rho,\bR)\right)\psi_{i}=\epsilon_{i}\psi_{i}, \quad i=1,2,...\,,
\end{equation}
where $\epsilon_{i}$ and $\psi_{i}$ denote the eigevalues and the corresponding eigenfunctions of the Kohn-Sham Hamiltonian, respectively, $\rho$ is the electron charge density of the non-interacting system, $\bR=\{\bR_{1}, \bR_{2}, \ldots, \bR_{N_{a}}\}$ is the collective representation for all nuclear positions in the system, and $V_{\text{eff}}(\rho,\bR)$ is the effective single-electron Kohn-Sham potential. In the present work, we limit our discussion to a non-periodic setting and spin-independent Hamiltonian. However, we note that all the ideas discussed subsequently can be generalized, in a straightforward manner, to periodic~\cite{Pask2005} or semi-periodic systems and spin-dependent Hamiltonians~\cite{Martin2004}.

The effective single-electron potential, $V_{\text{eff}}(\rho,\bR)$, in Eq.~\ref{eq:KSEig} is given by
\begin{equation} \label{eq:veff}
  V_{\text{eff}}(\rho,\bR)=V_{\text{xc}}(\rho)+V_{\text{H}}(\rho)+V_{\text{ext}}(\bx,\bR) \,.
\end{equation}
Here, $V_{\text{xc}}(\rho) = \frac{\delta E_{\text{xc}}}{\delta\rho}$ is the exchange-correlation potential and is defined as the variational derivative of the exchange-correlation energy, $E_{\text{xc}}$, with respect to $\rho$. Physically, $V_{\text{xc}}(\rho)$ is the mean-field potential that models the many-body interactions between electrons. In the present work, we have used the local density approximation (LDA)~\cite{Martin2004} for the exchange-correlation, specifically, the Ceperley-Alder~\cite{Ceperley80} form. The Hartree potential, $V_{\text{H}}(\rho)$, and the external potential, $V_{\text{ext}}(\bx,\bR)$, in Eq.~\ref{eq:veff}  are the classical electrostatic potentials corresponding to the electron charge density and nuclear charges, respectively, and are given by
\begin{equation} 
  V_{\text{H}}(\rho)=\int{\frac{\rho(\bx')}{|\bx-\bx'|}\dx'} \,,
\end{equation}
\begin{equation} \label{eq:vext}
  V_{\text{ext}}(\bx,\bR)=-\sum_{I=1}^{N_a}{\frac{Z_I}{|\bx-\bR_I|}} \,,
  \end{equation}
where $Z_I$ denotes the atomic number of the $I$th nucleus in the system.

We note that both the electrostatic potentials---Hartree ($V_{\text{H}}$) and external potential ($V_{\text{ext}}$)---are extended in real space. However, noting that the $\frac{1}{|\br|}$ kernel in these extended interactions is the Green's function of the Laplace operator, one can reformulate their evaluation as solutions of the Poisson problems, given by
\begin{subequations} \label{eq:Poisson}
  \begin{equation} \label{eq:Hartree_possion}
    -\frac{1}{4\pi}\nabla^2V_{\text{H}}(\bx)=\rho(\bx)\,,
  \end{equation}    
  \begin{equation} \label{eq:External_poisson}
    -\frac{1}{4\pi}\nabla^2V_{\text{ext}}(\bx,\bR)=b(\bx,\bR)\,.
  \end{equation}
\end{subequations}
In the above Eq.~\ref{eq:External_poisson}, we define $b(\bx,\bR)=-\sum_{I}^{N_a}{Z_I\widetilde{\delta}(\bx,\bR_{I})}$, where $\widetilde{\delta}(\bx,\bR_{I})$ is a Dirac distribution centered at $\bR_{I}$. We refer to previous works on finite element based DFT calculations ~\cite{Pask1999, Pask2005, Suryanarayana2010, Motamarri2012, Motamarri2013} for a comprehensive treatment of the local reformulation of electrostatic potentials into Poisson problems.

The electron charge density, the central quantity of interest in DFT, is given in terms of the eigenfunctions in Eq. ~\ref{eq:KSEig} as:
\begin{equation} \label{eq:Rho}
  \rho(\bx)=2\sum_{i}f(\epsilon_{i},\mu)|\psi_{i}(\bx)|^{2}\,,
\end{equation}
where $f(\epsilon,\mu)$ is the orbital occupancy function and $\mu$ is the Fermi energy. Typically, in DFT calculations the orbital occupancy function $f$ is chosen as the Fermi-Dirac distribution ~\cite{Kresse96,Goedecker99}, given by
\begin{equation} \label{eq:FermiDirac}
  f(\epsilon,\mu)=\frac{1}{1+\text{exp}(\frac{\epsilon-\mu}{k_{B}T})}\,,
\end{equation}
where $k_{B}$ denotes the Boltzman constant and $T$ is the temperature used for smearing the orbital occupancy function. The Fermi energy, $\mu$, is evaluated from the constraint on the total number of electrons $(N_e)$ in the system, given by
\begin{equation} \label{eq:Mu}
  \int{\rho(\bx)\dx}=2\sum_{i}f(\epsilon_{i},\mu)=N_e\,.
\end{equation}
The choice of a Fermi-Dirac distribution is made over a Heavyside function to avoid charge sloshing, wherein systems with degenerate energy levels at Fermi energy can exhibit large spatial deviation in electron charge density with SCF iterations on the account of different degenerate orbitals being occupied at different SCF iterations.

Finally, upon solving Eqs. ~\ref{eq:KSEig}, ~\ref{eq:Rho} and ~\ref{eq:Mu} self-consistently, the ground-state energy of the materials system is computed as
\begin{equation}
  E_{\text{tot}}=E_{\text{band}}+E_{\text{xc}}-\int{V_{\text{xc}}(\rho)\rho\dx}-\frac{1}{2}\int{\rho V_{\text{H}}(\rho)\dx}+E_{\text{ZZ}} \,,
\end{equation}
where $E_{\text{band}}$ is the band energy, given by
\begin{equation} \label{eq:bandEnergy}
  E_{\text{band}}=2\sum_i{f(\epsilon_i,\mu)\epsilon_i}\,,
\end{equation}
and $E_{\text{ZZ}}$ is the nuclear-nuclear repulsion, given by
\begin{equation} \label{eq:repulsiveEnergy}
  E_{\text{ZZ}}=\frac{1}{2}\sum_{\substack{I,J=1 \\ I\neq J}}^{N_a}{\frac{Z_IZ_J}{|\bR_I-\bR_J|}}\,.
\end{equation}

\section{Classical Finite Element Method}\label{sec:CFEM}
In this section, we briefly discuss the discretization of the Kohn-Sham eigenvalue problem using the classical finite element basis. In particular, we comment on the usefulness of higher-order spectral finite elements, employed in this work, which in conjunction with the reduced order Gauss-Lobatto-Legendre (GLL)  quadrature rule enables an efficient inversion of the overlap matrix of the classical finite element basis functions. 

\subsection{Classical finite element discretization} \label{sec:CFEDiscrete}
In the finite element method, the spatial domain of interest is discretized into subdomains called finite elements using a finite element mesh. The finite element basis is constructed from piecewise polynomial functions that have a compact support on the finite elements, thus rendering locality to these basis functions. We note that there is an abundance of choice in terms of the form and order of the polynomial functions that can be used in constructing the finite element basis, and we refer to~\cite{Hughes2012, Bathe1996} for a comprehensive discourse on the subject. Henceforth, we refer to the standard notion of finite element basis as the classical finite element basis in order to differentiate from the proposed enriched finite element basis in section~\ref{sec:EFEM}, and refer to the corresponding discrete formulation as the classical finite element discretization.

Let $X_h$ denote the finite element subspace of dimension $n_h$ constructed from a finite element mesh with a characteristic mesh-size $h$. Let $\psi_i^h$ and $\phi^h$ denote the classical finite element discretized fields corresponding to the Kohn-Sham orbitals and the electrostatic potential (generically representing both Hartree and external potential), respectively, that are expressed as
\begin{subequations} \label{eq:CFEM}
  \begin{equation} \label{eq:CFEMDiscretePsi}
    \psi_{i}^{h}(\bx) =\sum_{j=1}^{n_h}{N_j^C(\bx)\psi_{i,j}^C} \qquad i=1,2, \ldots\,,
  \end{equation}
  \begin{equation} \label{eq:CFEMDiscretePhi}
    \phi^{h}(\bx) =\sum_{j=1}^{n_h}{N_j^C(\bx)\phi_j^C}\,.
  \end{equation}
\end{subequations}
The superscript $C$, in the above expressions and elsewhere in the article, is used to indicate the discretization based on classical finite element basis. Here $N_j^C:1\leq j \leq n_h$ denote the classical finite element basis functions spanning $X_h$, and $\psi_{i,j}^C$ and $\phi_j^C$ are the coefficients corresponding to $j$-th basis function ($N_j^C$) in the expansion of the $i$-th Kohn-Sham orbital and electrostatic potential, respectively. 

Using the classical finite element discretization in Eq.~\ref{eq:CFEMDiscretePsi}, the Kohn-Sham eigenvalue problem in Eq. ~\ref{eq:KSEig} reduces to the following discrete form,
\begin{equation} \label{eq:DiscreteKS}
  \mathbf{H}^C\Psi_i^C=\epsilon_i^C\mathbf{M}^C\Psi_i^C\,,
\end{equation}
where $\mathbf{H}^C$ denotes the discrete Kohn-Sham Hamiltonian, $\mathbf{M}^C$ denotes the overlap matrix of the classical finite element basis, $\epsilon_i^C$ denotes the $i$-th discrete Kohn-Sham eigenvalue, and $\Psi_i^C$ denotes the corresponding eigenvector containing the expansion coefficients $\psi_{i,j}^C$. For a non-periodic problem defined on a domain $\Omega$ with homogeneous Dirichlet boundary conditions, the discrete Hamiltonian matrix $H_{jk}^C$ is given by
\begin{equation} \label{eq:Hjk}
  \begin{split}
  H_{jk}^C=\frac{1}{2}\int_{\Omega}{\nabla N_j^C(\bx).\nabla N_k^C(\bx)\dx} ~+ \\
	    \int_{\Omega}{V_{\text{eff}}^{h}(\bx,\bR)N_j^C(\bx)N_k^C(\bx)\dx}\,.
  \end{split}
\end{equation}
Although the above expression is for a non-periodic problem, it can be easily extended to a periodic problem on a unit cell using the Bloch theorem ~\cite{Pask2005}. 
We note that owing to the non-orthogonality of the classical finite element basis, the overlap matrix $\mathbf{M}^C$, defined by $M_{jk}^C=\int_{\Omega}{N_j^C(\bx)N_k^C(\bx)\dx}$, is not an identity matrix, thereby, resulting in a generalized eigenvalue problem. However, utilizing the symmetric positive definiteness, and hence the invertibility of $\mathbf{M}^C$, we can transform the generalized eigenvalue problem in Eq. ~\ref{eq:DiscreteKS} to a standard eigenvalue problem, given by
\begin{equation} \label{eq:SEPCFEM}
  {(\mathbf{M}^C)}^{-1}\mathbf{H}^C\Psi_i^C=\epsilon_i^C\Psi_i^C\,.
\end{equation}
We remark that this transformation of the generalized eigenvalue problem to a standard eigenvalue problem is essential for the use of Chebyshev polynomial based acceleration technique to compute the occupied eigenspace (to be discussed in the Section ~\ref{sec:ChFSI}). Further, we note that this transformation to a standard eigenvalue problem relies on computationally efficient methods for computing ${(\mathbf{M}^C)}^{-1}$, which forms the basis for our use of spectral finite elements along with Gauss-Lobatto-Legendre quadrature rule, as will be discussed in Section ~\ref{sec:Spectral}. 

Turning to the Poisson problems in Eq. ~\ref{eq:Poisson}, and using the discretization in Eq. ~\ref{eq:CFEMDiscretePhi}, we obtain the following system of linear equations,
\begin{equation} \label{eq:discretePoissonCFEM}
  \mathbf{A}^C\Phi^C=\mathbf{b}^C \,,
\end{equation}
where $\mathbf{A}^C$ represents the Laplace operator discretized in the classical finite element basis that is given by
\begin{equation}
  A_{jk}^C=\frac{1}{4\pi}\int_{\Omega}{\nabla N_j^C(\bx).\nabla N_k^C(\bx)\dx} \,,
\end{equation}
$\Phi^C$ is the electrostatic potential vector containing the expansion coefficients $\phi_j^C$, and $\mathbf{b}^C$, the forcing vector, is given by
\begin{equation} \label{eq:ForceCFEM}
  b_i^C=\int_{\Omega}{g(\bx)N_i^C(\bx)\dx}\,,
\end{equation}
where $g(\bx)=\rho(\bx)$ or $g(\bx)=b(\bx,\bR)$ for the Hartree and external potential, respectively.

\subsection{Spectral finite elements} \label{sec:Spectral}
As opposed to conventional classical finite element basis, which is typically constructed from a tensor product of Lagrange polynomials interpolated through equidistant nodal points in an element, spectral finite element basis employ a distribution of nodes obtained from the roots of the derivative of Legendre polynomials or the Chebyshev polynomials~\cite{Boyd2013}. In the present work, we employ the Gauss-Lobatto-Legendre node distribution, where the nodes are located at the roots of the derivative of the Legendre polynomial and the boundary points. The resulting spectral finite element basis has been shown to provide better conditioning with increasing polynomial degree~\cite{Boyd2013} and has been effective for electronic structure calculations using higher-order finite element discretization~\cite{Motamarri2012, Motamarri2013}. However, the major advantage of this spectral finite element basis is realized when it is used in conjunction with Gauss-Lobatto-Legendre (GLL) quadrature rule~\cite{Canuto2007} for evaluation of the integrals arising in the overlap matrix, wherein the quadrature points are coincident with the nodal points in the spectral finite element. Such a combination renders the overlap matrix $\mathbf{M}^C$ in the discrete Kohn-Sham eigenvalue problem diagonal, thereby making the transformation of the generalized eigenvalue problem in Eq. ~\ref{eq:DiscreteKS} to the standard eigenvalue problem in Eq. ~\ref{eq:SEPCFEM} to be trivial. We note that while an $n$ point rule in the conventional Gauss quadrature rule integrates polynomials exactly up to degree $2n-1$, an $n$ point GLL quadrature rule integrates polynomials exactly only up to degree $2n-3$. Thus, we employ the GLL quadrature rule only in the construction of $\mathbf{M}^C$, while the more accurate Gauss quadrature rule is used for all other integrals featuring in the Kohn-Sham eigenvalue problem as well as the Poisson problems for the electrostatic potentials. We refer to Motamarri et. al ~\cite{Motamarri2013} for a discussion on the accuracy and sufficiency of GLL quadrature in the evaluation of overlap matrix $\mathbf{M}^C$. Since we employ spectral finite elements all throughout the present work, any reference to classical finite elements, henceforth, corresponds to spectral finite elements.

\section{Enriched Finite Element Method} \label{sec:EFEM}
In this section, we first discuss the proposed enriched finite element discretization for the Kohn-Sham eigenvalue problem. Then, we present various numerical and algorithmic strategies for efficient use of the enriched finite element basis.

\subsection{Enriched finite element discretization} \label{sec:EFEMDiscretize}
In enriched finite element discretization we augment the classical finite element basis by appending additional atom-centered basis functions called enrichment functions. We write the enriched finite element discretization for the Kohn-Sham orbitals, $\psi_i^h$, and the electrostatic potentials (both Hartree and external potential), $\phi^h$, as follows:
\begin{subequations} \label{eq:EFEM}
  \begin{equation} \label{eq:EFEMPsi}
    \psi_i^h(\bx)=\underbrace{\sum_{j=1}^{n_h}{N_j^C(\bx)\psi_{i,j}^C}}_{\text{Classical}}+
		  \underbrace{\sum_{I=1}^{N_a}\sum_{k=1}^{n_I}{N_{k,I}^{E,\psi}(\bx,\bR_I)\psi_{i,k,I}^E}}_{\text{Enriched}}\,,
  \end{equation}

  \begin{equation} \label{eq:EFEMPhi}
    \phi^h(\bx)=\underbrace{\sum_{j=1}^{n_h}{N_j^C(\bx)\phi_j^C}}_{\text{Classical}}+
		  \underbrace{\sum_{I=1}^{N_a}{N_I^{E,\phi}(\bx,\bR_I)\phi_I^E}}_{\text{Enriched}}\,.
  \end{equation}
\end{subequations}
 In the above expressions, the superscripts $C$ and $E$ are used to distinguish between classical and enriched components, respectively. As with the classical finite element discretization, $N_j^C$ denotes the $j$-th classical finite element basis, and $\psi_{i,j}^C$ and $\phi_j^C$ are the expansion coefficients corresponding to $N_j^C$ for the $i$-th Kohn-Sham orbital and the electrostatic potential, respectively. In addition, we have the enrichment functions $N_{k,I}^{E,\psi}$ and $N_I^{E,\phi}$ for the Kohn-Sham orbitals and the electrostatic potentials, respectively, each centered on atom $I$ located at $\bR_I$. $\psi_{i,k,I}^E$ denotes the expansion coefficient correspoding to $N_{k,I}^{E,\psi}$ for the $i$-th Kohn-Sham orbital, and $\phi_I^E$ denotes the expansion coefficient corresponding to $N_I^{E,\phi}$ for the electrostatic potential. The enrichment functions, $N_{k,I}^{E,\psi}$ and $N_I^{E,\phi}$, are the atom-centered numerical solutions to the Kohn-Sham orbitals and electrostatic potentials, respectively, for the atom type (chemical element) at $\bR_I$. The index $I$ runs over all the atoms, $N_a$, in the materials system, and the index $k$ in Eq. ~\ref{eq:EFEMPsi} runs over the number of atomic Kohn-Sham orbitals, $n_I$, we would want to include for the atom $I$. Typically, we include all the occupied and a few lowest unoccupied atomic orbitals for a given atom $I$. We note that although we have represented the enrichment functions for both Hartree and external potential as $N_I^{E,\phi}$, they differ based on the electrostatic potential that is being discretized.

We now discuss the procedure to generate the enrichment functions. As aforementioned, the enrichment functions are chosen as the solutions to the Kohn-Sham orbitals and electrostatic potentials for any given single atom. Under the assumption of equal fractional occupancy for degenrate orbitals, the charge density for a single atom is radially symmetric, which in turn, results in radially symmetric $V_{\text{xc}}(\rho)$ and $V_{\text{H}}(\rho)$. Thus, rewriting the Eqs. ~\ref{eq:KSEig} and ~\ref{eq:Poisson} in spherical coordinates and using separation of variables, we obtain the following radial equations for any single atom with the atom type denoted by a superscript $S$:
\begin{subequations} \label{eq:1DRadial}
  \begin{equation} \label{eq:RadialPoisson}
    -\frac{1}{4\pi}\frac{1}{r^2}\frac{d}{dr}\bigl(r^2\frac{d}{dr}\bigr)\phi^S(r)=g^S(r)\,,
  \end{equation}
  \begin{equation} \label{eq:RadialKS}
    \Big[-\frac{1}{2}\frac{1}{r^2}\frac{d}{dr}\bigl(r^2\frac{d}{dr}\big)+\frac{l(l+1)}{r^2}+V^S_{\text{eff}}(r)\Big]R_{nl}^S(r)=\epsilon_{nl}^SR_{nl}^S(r)\,,
  \end{equation}
  \begin{equation} \label{eq:RadialRho}
    \rho^S(r)=2\sum_n\sum_l \frac{2l+1}{4\pi}f(\epsilon^S_{nl},\mu^S){\bigl(R_{nl}^S(r)\bigr)}^2\,.
  \end{equation}
\end{subequations}
In Eq. ~\ref{eq:RadialPoisson}, $\phi^S(r)$ denotes either the Hartree or the external potential; $g^S(r)$ denotes the charge density $\rho^S(r)$ or the nuclear charge $b^S(r)=Z_S\widetilde{\delta}(0)$ with $Z_s$ denoting the atomic number, depending on whether $\phi^S(r)$ represents the Hartree or the external potential, respectively. In Eq. ~\ref{eq:RadialKS}, $R_{nl}^S(r)$ represents the radial part of the Kohn-Sham orbital corresponding to the principal quantum number $n$ and azimuthal quantum number $l$. Equations in~\ref{eq:1DRadial} are solved self-consistently until convergence in $\rho^S(r)$ is achieved. We note that these radial equations can be solved inexpensively using a 1D classical finite element mesh comprising of, typically, $1000-5000$ basis functions. Moreover, the radial atomic solutions can be pre-computed for all atom types spanning the periodic table and stored for later use in constructing the enrichment functions. 

Having evaluated the radial part $R_{nl}^S(r)$, the full Kohn-Sham orbital is obtained by multiplying it with spherical harmonics as follows
\begin{equation} \label{eq:KSOrbitalSplit}
  \psi_{nlm}^S(r,\beta,\gamma)=R_{nl}^S(r)Y_{lm}(\beta,\gamma)\,,
\end{equation}
where $Y_{lm}(\beta,\gamma)$ denotes the real form of spherical harmonics for the pair of azimuthal quantum number $l$ and magnetic quantum number $m$, and $\beta$ and $\gamma$ represent the polar and azimuthal angles, respectively. Using the above atomic solutions, we write the orbital enrichment function $N_{k,I}^{E,\psi}$ centered at atom $I$ of atom type $S$ as
\begin{equation} \label{eq:EnrichedPsi}
  N_{k,I}^{E,\psi}(\bx, \bR_I)=\psi_{nlm}^S(|\bx-\bR_I|,\beta_{\bR_I},\gamma_{\bR_I})\,,
\end{equation}
where the index $k$ represents a specific choice of $n$, $l$ and $m$, and $\beta_{\bR_I}$ and $\gamma_{\bR_I}$ are the polar and azimuthal angles, respectively, for the point $\bx$ with $\bR_I$ as the origin. Similarly, we define the electrostatic enrichment function $N_I^{E,\phi}(\bx)$ centered at atom $I$ of atom type $S$ as
\begin{equation} \label{eq:EnrichedPhi}
  N_I^{E,\phi}(\bx, \bR_I)=\phi^S(|\bx-\bR_I|)\,.
\end{equation}
Henceforth in the paper, to make our notation of the enrichment functions more succinct, we combine the indices $k$ and $I$ into a single index  denoted by $\alpha$ for the orbital enrichment functions and their coefficients, and drop the argument $\bR_I$ in the enrichment functions. Furthermore, we define $n_E^{\psi} = N_a \times n_I$ to denote the total number of enrichment functions in the materials system used for discretization of any Kohn-Sham orbital $\psi_i$.

Discretizing the Kohn-Sham eigenvalue problem in the enriched finite element basis, we obtain a standard eigenvalue equation analogous to its classical counterpart (Eq. ~\ref{eq:SEPCFEM}), and is given by
\begin{equation} \label{eq:SEPEFEM}
  {(\mathbf{M}^E)}^{-1}\mathbf{H}^E\Psi_i^E=\epsilon_i^E\Psi_i^E\,,
\end{equation}
where $\mathbf{H}^E$ and $\mathbf{M}^E$ are the discrete Kohn-Sham Hamiltonian matrix and overlap matrix in the enriched finite element basis, $\epsilon_i^E$ denotes the $i$-th discrete Kohn-Sham eigenvalue and $\Psi_i^E$ denotes the corresponding eigenvector containing the expansion coefficients $\psi_{i,j}^C$ and $\psi_{i,\alpha}^E$ (defined in Eq. ~\ref{eq:EFEMPsi}). Both $\mathbf{H}^E$ and $\mathbf{M}^E$ matrices have a $2\times2$ block structure, given by
\begin{equation} \label{eq:H2x2}
\mathbf{H}^E = 
\left[\begin{array}{M{2.6cm}|M{1.2cm}}
  \\  $\mathbf{H^{cc}}$  &	$\mathbf{(H^{ec})}^T$ \\ \\\hline
    $\mathbf{H^{ec}}$  &	$\mathbf{H^{ee}}$
\end{array}\right]
\end{equation}

\begin{equation} \label{eq:M2x2}
\mathbf{M}^E = 
\left[\begin{array}{M{2.6cm}|M{1.2cm}}
  \\  $\mathbf{M^{cc}}$  &	$\mathbf{(M^{ec})}^T$ \\ \\\hline
    $\mathbf{M^{ec}}$  &	$\mathbf{M^{ee}}$
\end{array}\right]
\end{equation}
where $\mathbf{H^{cc}}$ and $\mathbf{M^{cc}}$ are the classical-classical blocks which comprise of matrix elements involving only the classical finite element basis functions and are same as the $\mathbf{H}^C$ and $\mathbf{M}^C$ matrices appearing in Eq. ~\ref{eq:DiscreteKS}, respectively; $\mathbf{H^{ec}}$ and $\mathbf{M^{ec}}$ are the enriched-classical blocks containing the cross-term matrix elements involving both classical finite element basis functions and enrichment functions; and $\mathbf{H^{ee}}$ and $\mathbf{M^{ee}}$ are the enriched-enriched blocks comprising of matrix elements involving only the enrichment functions. Each of these blocks are given by
\begin{subequations} \label{eq:HBlocks}
  \begin{equation} \label{eq:Hcc}
    \begin{split}
    H^{cc}_{jk}=\frac{1}{2}\int_{\Omega}{\nabla N_j^C(\bx).\nabla N_k^C(\bx)\dx} ~+ \\
		      \int_{\Omega}{V_{\text{eff}}^{h}(\bx,\bR)N_j^C(\bx) N_k^C(\bx)\dx}\,,
    \end{split}
  \end{equation}
  \begin{equation} \label{eq:Hec}
    \begin{split}
      H^{ec}_{\alpha j}=\frac{1}{2}\int_{\Omega}{\nabla N_{\alpha}^{E,\psi}(\bx).\nabla N_j^C(\bx)\dx} ~+ \\
		      \int_{\Omega}{V_{\text{eff}}^{h}(\bx,\bR)N_{\alpha}^{E,\psi}(\bx)N_j^C(\bx)\dx}\,,
    \end{split}
  \end{equation}
  \begin{equation} \label{eq:Hee}
    \begin{split}
      H^{ee}_{\alpha\beta}=\frac{1}{2}\int_{\Omega}{\nabla N_{\alpha}^{E,\psi}(\bx).\nabla N_{\beta}^{E,\psi}(\bx)\dx} ~+ \\ 
			  \int_{\Omega}{V_{\text{eff}}^{h}(\bx,\bR)N_{\alpha}^{E,\psi}(\bx) N_{\beta}^{E,\psi}(\bx)\dx}\,;
    \end{split}
  \end{equation}
\end{subequations}

\begin{subequations} \label{eq:MBlocks}
  \begin{equation} \label{eq:Mcc}
    M^{cc}_{jk}=\int_{\Omega}{N_j^C(\bx) N_k^C(\bx)\dx}\,,
  \end{equation}
  \begin{equation} \label{eq:Mec}
    M^{ec}_{\alpha j}=\int_{\Omega}{N_{\alpha}^{E,\psi}(\bx) N_j^C(\bx)\dx}\,,
  \end{equation}
  \begin{equation} \label{eq:Mee}
    M^{ee}_{\alpha\beta}=\int_{\Omega}{N_{\alpha}^{E,\psi}(\bx) N_{\beta}^{E,\psi}(\bx)\dx}\,,
  \end{equation}
\end{subequations}
where $j,k=1,2,\ldots,n_h$ and $\alpha,\beta=1,2,\ldots,n_E^{\psi}$.

Discretizing the Poisson problems (Eq. ~\ref{eq:Poisson}) in the enriched finite element basis, we obtain a system of linear equations analogous to its classical counterpart (Eq. ~\ref{eq:discretePoissonCFEM}), and is given by
\begin{equation} \label{eq:discretePoissonEFEM}
  \mathbf{A}^E\Phi^E=\mathbf{b}^E\,,
\end{equation}
where $\mathbf{A}^E$ represents the discrete Laplace operator in the enriched finite element basis, and $\Phi^E$ is the electrostatic potential vector containing the expansion coefficients $\phi_j^C$ and $\phi_I^E$ (defined in Eq. ~\ref{eq:EFEMPhi}). 
Similar to $\mathbf{H}^E$ and $\mathbf{M}^E$, the matrix $\mathbf{A}^E$ also assumes a $2\times2$ block structure containing classical-classical, enriched-classical and enriched-enriched blocks, given by
\begin{equation} \label{eq:A2x2}
\mathbf{A}^E = 
\left[\begin{array}{M{2.6cm}|M{1.2cm}}
  \\  $\mathbf{A^{cc}}$  &	$\mathbf{(A^{ec})}^T$ \\ \\\hline
    $\mathbf{A^{ec}}$  &	$\mathbf{A^{ee}}$
\end{array}\right]
\end{equation}
with the individual blocks defined as
\begin{subequations} \label{eq:ABlocks}
  \begin{equation} \label{eq:Acc}
    A^{cc}_{jk}=\int_{\Omega}{\nabla N_j^C(\bx) . \nabla N_k^C(\bx)\dx}\,,
  \end{equation}
  \begin{equation} \label{eq:Aec}
    A^{ec}_{Ij}=\int_{\Omega}{\nabla N_I^{E,\phi}(\bx) . \nabla N_j^C(\bx)\dx}\,,
  \end{equation}
  \begin{equation} \label{eq:Aee}
    A^{ee}_{IJ}=\int_{\Omega}{\nabla N_I^{E,\phi}(\bx) . \nabla N_J^{E,\phi}(\bx)\dx}\,,
  \end{equation}
\end{subequations}
where $j,k=1,2,\ldots,n_h$; and $I,J=1,2,\ldots,N_a$.

The forcing vector $\mathbf{b}^E$, is also analogous to its classical counterpart, and is defined as the composite vector
\begin{equation} \label{eq:b2x1}
\mathbf{b}^E = 
\left[\begin{array}{M{1.2cm}}
  \\  $\mathbf{b^{c}}$ \\ \\ \hline
    $\mathbf{b^{e}}$  
\end{array}\right]
\end{equation}
where $\mathbf{b^{c}}$ is the classical part of $\mathbf{b}^E$ and is same as $\mathbf{b}^{C}$ (defined in Eq. ~\ref{eq:ForceCFEM}). $\mathbf{b^{e}}$ is the enrichment part of $\mathbf{b}^E$ and is given by
\begin{equation}
  b_I^e=\int_{\Omega}{g(\bx)N_I^{E,\phi}(\bx)\dx}\,,
\end{equation}
where $g(\bx)=\rho(\bx)$ or $g(\bx)=b(\bx,\bR)$ for the Hartree and external potential, respectively, and $I=1,2,\ldots,N_a$.

The key idea behind augmenting the classical finite element basis with these enrichment functions is that in a multi-atom materials system, the enrichment functions, being solutions to single atom Kohn-Sham orbitals and electrostatic potentials, can effectively capture the sharp variations in the orbitals and the electrostatic potentials close to an atom, thereby eliminating the need for a refined classical finite element mesh close to an atom. Loosely speaking, the role of the classical finite element basis is now to capture the deviation of an electronic field in a materials system from that of superposition of atomic solutions for the same field. Since these deviations are much smoother in nature compared to the actual field, we can use a coarse classical finite element mesh to accurately approximate them. As will be discussed in Section~\ref{sec:ChFSI}, the use of a coarse classical finite element mesh results in two-fold advantage: (i) a reduction in the total degrees of freedom, and (ii) a reduction in the polynomial degree of the Chebyshev filter required to compute the occupied Kohn-Sham eigenspace.

\subsection{Conditioning of the enriched finite element basis} \label{sec:Mollifier}
The enrichment functions, being solutions to the Kohn-Sham orbitals and electrostatic potentials for a single atom, have smooth tails away from their atomic cores. These smooth tails can cause linear dependency between the enrichment functions and the classical finite element basis, thereby resulting in an ill-conditioned basis. We avoid such ill-conditioning by multiplying the enrichment functions with a smooth radial cutoff function, which generates a compact support for each enrichment function. In the present work, we employ the smooth cutoff function given by
\begin{equation} \label{eq:Moll1}
  h(\widetilde{r})=\frac{u(\widetilde{r})}{u(\widetilde{r})+u(1-\widetilde{r})}\,,
\end{equation}
where $u(\widetilde{r})$ is defined as
  \[ u(\widetilde{r}) = \begin{cases}
	      e^{-\frac{1}{\widetilde{r}}} & \widetilde{r} > 0 \\
	      0 & \widetilde{r} \leq 0
	    \end{cases}
  \]
and $\widetilde{r}=1-\frac{t(r-r_0)}{r_0}$ is a linear transformation of the variable $r$, which offers $h(\widetilde{r})$ the following properties
  \[ \begin{cases}
    h(\widetilde{r})=1 & 0 \leq r < r_0 \\
	0 \leq h(\widetilde{r}) < 1 & r_0 < r \leq r_0 + \frac{r_0}{t}\\
	h(\widetilde{r})= 0 & r > r_0 + \frac{r_0}{t}  \,\,.
      \end{cases}
  \]
We multiply the radial part of each enrichment function, $N_{\alpha}^{E,\psi}(\bx)$ or $N_I^{E,\phi}$, with $h(\widetilde{r})$ to smoothly truncate them to zero. In the above expression, the parameter $r_0$ is called the cutoff radius, beyond which the truncation begins, and $t$ controls the width of the transition. In the present work, we employ different values of $r_0$ for different enrichment functions. In particular, for an orbital enrichment function, the value of $r_0$ is chosen to be the farthest turning point (extremum) in the radial part of the corresponding atomic orbital. One exception to this rule is the monotonically decreasing $1s$ radial function, $R_{10}(r)$, for which the $r_0$ is chosen such that $|\int_0^{r_0}{\left(h(\widetilde{r})R_{10}(r)\right)}^2\;dr - 1| < 10^{-6}$, i.e., the density arising out of the truncated $R_{10}(r)$ must integrate to within $10^{-6}$ of unity. The maximum of the set of $r_0$'s corresponding to orbital enrichment functions of a given atom is selected as the cutoff radius for the electrostatic enrichment functions of the atom. We use $t \in [0.5,1]$ to avoid sharp truncation of the enrichment functions, which may otherwise require a very high density of quadrature points in the transition region in order to accurately compute any integrals involving the gradients of these truncated enrichment functions. Figure ~\ref{fig:Mollifier} presents a schematic of the radial part of the truncated atomic orbital.
\begin{figure} [!] \label{fig:Mollifier} 
  \centering
  \includegraphics[width=1.0\columnwidth]{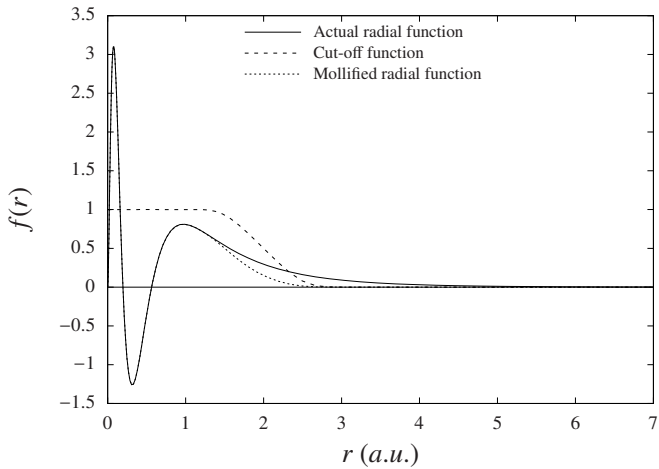}
  \caption{Schematic of truncated atomic orbital (radial part).}
  \label{fig:Mollifier}
\end{figure}
Henceforth, enrichment functions, $N_{\alpha}^{E,\psi}(\bx)$ or $N_I^{E,\phi}(\bx)$, are assumed to be truncated with the aforementioned smooth cutoff function. We remark that, in addition to improving the conditioning of the basis, the truncation renders locality to the enrichment functions, which in turn renders sparsity to the discrete Hamiltonian, Laplacian and overlap matrices. 

We note that several prior efforts have been made towards the generation of compactly supported (finite-range) atom-centered orbitals by employing different forms of confining potentials in the atomic Kohn-Sham equation, ranging from hard-wall potential ~\cite{Sankey1989} to polynomial ~\cite{Horsfield1997, Porezag1995} to smooth exponential potential ~\cite{Blum2009}. Other efforts ~\cite{Junquera2001, Anglada2002} were made to variationally optimize the parameters in the confining potential to strike a good balance between the locality and accuracy of the resultant basis. In our view all these approaches can be adapted as an alternative to our approach of using smooth cutoff function. 
\subsection{Adaptive quadrature rule}
We note that sharp gradients in regions close to atomic centers and cusps at atomic centers are characteristics of enrichment functions. Therefore, in order to accurately compute any integral involving an enrichment function, we need a high quadrature density near the atomic centers, while a lower quadrature density may suffice in regions farther away from atomic centers. To this end, we employ an adaptive refinement of the quadrature grid on each finite element based on the characteristics of the enrichment functions. Specifically, we follow a divide and conquer strategy proposed in previous efforts~\cite{Mousavi2012, Pieper1999, Berntsen1991}, and outline here the main idea and specifics of our implementation for hexahedral finite elements employed in this work. For any given finite element, we begin by assigning it to be the parent element $\Omega^p$. Further, we consider a trial function $f(\bx)$, an $n$-point Gauss quadrature rule, the $8$ child elements $\left(\left\{\Omega^c_i\right\}\right)_{i=1}^{8}$ that are obtained by sub-dividing $\Omega^p$, a fixed tolerance $\tau$, and an empty list labelled \textit{points}. Next, we evaluate $I^p=\int_{\Omega^p}{f(\bx)\dx}$ and $I^c_i=\int_{\Omega^c_i}{f(\bx)\dx}$ for $i=1,2,\ldots,8$, using their respective $n$-point Gauss quadrature rules. If the base condition, $|I^p-\sum_{i=1}^{8}{I^c_i}|<\tau$, is satisfied, we append the Gauss quadrature points and weights of the parent element to the list \textit{points} and terminate the algorithm. Otherwise, we treat each of the child elements as a parent element, and recursively sub-divide them until the base condition is satisfied. Whenever the base condition is satisfied, the Gauss quadrature points and weights corresponding to the parent element at the current recursion level are appended to the list \textit{points}. Finally, the list \textit{points} represents the quadrature points and weights for the given element. We repeat this process for each element present in the finite element mesh. Typically, instead of using a single trial function $f(\bx)$, we use $n_t$ such trial functions, $\left\{f_{\nu}(\bx)\right\}_{\nu=1}^{n_t}$, which requires $n_t$ base conditions corresponding to each $f_{\nu}(\bx)$ to be satisfied.  

In the present work, we choose the following four trial functions to build the adaptive quadrature rule:
\begin{subequations} \label{eq:QuadTrialFunctions}
  \begin{equation}
    f_1(\bx)=\sum_{I=1}^{N_a}{\big(N_I^{E,\phi}(\bx)\big)}^2\,,
  \end{equation}
  \begin{equation}
    f_2(\bx)=\sum_{I=1}^{N_a}{|\nabla\big(N_I^{E,\phi}(\bx))|}^2\,,
  \end{equation}
  \begin{equation}
    f_3(\bx)=\sum_{\alpha=1}^{n_E^{\psi}}{\big(N_{\alpha}^{E,\psi}(\bx)\big)}^2\,,
  \end{equation}
  \begin{equation}
    f_4(\bx)=\sum_{\alpha=1}^{n_E^{\psi}}{|\nabla\big(N_{\alpha}^{\psi}(\bx))|}^2\,.
  \end{equation}
\end{subequations}

Although we have labeled just two trial functions, $f_1(\bx)$ and $f_2(\bx)$, defined by the electrostatic enrichment functions, these correspond to four trial functions---two each for enrichment functions corresponding to the Hartree potential and the external potential. We remark that the aforementioned adaptive quadrature construction is performed only on the finite elements which are within the compact support of the enrichment functions. Since only a small fraction of the total elements are within the compact support of any enrichment function, the adaptive quadrature construction is computationally inexpensive. Further, once constructed, the adaptive quadrature list remains fixed for a given finite element mesh, and only needs to be updated if the finite element mesh changes during the course of the calculation.

We now turn towards determining an economical choice for the tolerance parameter, $\tau$, as a loose tolerance may result in an inadequate quadrature grid whereas an extremely tight tolerance will be computationally inefficient. In the present work, we employ the following heuristic to choose $\tau$. For each atom type $S$ of atomic number $Z_S$ in the materials system, we obtain the atomic ground-state charge density, $\rho^S(r)$, its corresponding Hartree potential, $\phi^S_{\text{H}}(r)$, and the atomic external potential, $\phi^S_{\text{ext}}(r)$,  by solving the the radial Kohn-Sham equations in Eqs. ~\ref{eq:1DRadial}. Next, we evaluate the following two integrals $E^{S,1D}_1=\frac{1}{2}\int {4 \pi r^2 \rho^S(r)\phi^S_{\text{H}}(r)\,dr}$ and $E^{S,1D}_2=\int {4 \pi r^2 \rho^S(r)\phi^S_{\text{ext}}(r)\,dr}$, which correspond to the electrostatic interaction energies. We then construct a coarse 3D finite element mesh with atom $S$ at the origin. In order to determine a judicious choice for $\tau^S$ corresponding to atom type $S$, we set its initial value as $\tau^S=0.1$ and enter an iterative loop. For the current iterate of $\tau^S$, we evaluate the 3D counterparts of $E^{S,1D}_1$ and $E^{S,1D}_2$, namely, $E^{S,3D}_1$ and $E^{S,3D}_2$, respectively, using the aforementioned adaptive quadrature rule. If the convergence criteria, $|E^{S,1D}_1-E^{S,3D}_1| < \gamma$ and $|E^{S,1D}_2-E^{S,3D}_2| < \gamma$, are satisfied for a pre-determined $\gamma$, we terminate the loop with the current value of $\tau^S$. Else, the loop is repeated with $\tau^S$ set to half of its current value, until the above convergence criteria are met. We use the minimum of all such $\tau^S$ corresponding to the various atom types in the materials system as our $\tau$ for construction of the adaptive quadrature grid for the materials system calculation. In all our calculations, we have used $\gamma = 0.1\,$mHa so as to ensure that the quadrature errors are an order lower than the desired discretization error $(\sim1\,$mHa) that we are aiming in the ground-state energy per atom. We note that the above procedure to determine $\tau$, is independent of the choice of 3D finite element mesh. Moreover, the $\tau^S$ for each $S$ can be precomputed and stored for later use.

\subsection{Inverse of overlap matrix}
We now discuss a computationally efficient way of evaluating the inverse of the overlap matrix, $\mathbf{M}^E$, defined in Eq. ~\ref{eq:M2x2}, which is vital to the transformation of the generalized Kohn-Sham eigenvalue problem to a standard eigenvalue problem, and the subsequent use of Chebyshev polynomial based acceleration technique to compute the occupied eigenstates as will be discussed in Section~\ref{sec:ChFSI}. We make use of the block-wise matrix inversion theorem ~\cite{Zhang2006} (also known as Banachiewicz inversion formula), to obtain the following $2\times2$ block structure for ${(\mathbf{M}^E)}^{-1}$,\
\begin{equation}
{(\mathbf{M}^E)}^{-1}= 
\left[\begin{array}{M{3.0cm}|M{1.5cm}}
  \\  $${(\mathbf{M^{cc}})}^{-1}+\mathbf{L^T}\mathbf{S}^{-1}\mathbf{L}$$  &	$-\mathbf{L^T}\mathbf{S}^{-1}$ \\ \\\hline
    -$\mathbf{S}^{-1}\mathbf{L}$  &	$\mathbf{S}^{-1}$
\end{array}\right]
\end{equation}
where $\mathbf{L}=\mathbf{M^{ec}}{(\mathbf{M^{cc}})}^{-1}$, and $\mathbf{S}=\mathbf{M^{ee}}-\mathbf{M^{ec}}{(\mathbf{M^{cc})}^{-1}}{\mathbf{(M^{ec})}^T}$.
Assuming that the enriched finite element basis is not ill-conditioned, we note that the overlap matrix $\mathbf{M}^E$ is positive definite, and, hence invertible. Further, $\mathbf{M^{cc}}$ being the overlap matrix of the classical finite element basis functions, is also positive definite, and hence invertible. Subsequently, the positive definiteness, and hence invertibility, of  $\mathbf{S}$ can be ascertained by noting that it is the Schur complement ~\cite{Zhang2006} of $\mathbf{M^{cc}}$ in $\mathbf{M}^E$. We note that the above expression for ${(\mathbf{M}^E)}^{-1}$ contains two matrix inverses, ${(\mathbf{M^{cc}})}^{-1}$ and $\mathbf{S^{-1}}$. As mentioned in Section ~\ref{sec:Spectral}, the matrix $\mathbf{M^{cc}}$ is rendered diagonal through the use of spectral finite elements along with Gauss-Lobatto-Legendre quadrature rule, which makes the evaluation of ${(\mathbf{M^{cc}})}^{-1}$ trivial. Regarding the evaluation of $\mathbf{S}^{-1}$, we note that $\mathbf{S}$ is a small matrix of the size of $n_E^{\psi}\times n_E^{\psi}$, where $n_E^{\psi}$ is typically of the same order as the number of electrons in the system. Thus, $\mathbf{S}$ can be easily inverted through the use of direct solvers.

Further, we note that although the overlap matrix is sparse, its inverse is a dense matrix. However, the constituent matrices present in the $2\times2$ block structure of ${(\mathbf{M}^E)}^{-1}$ are either sparse or much smaller in size compared to ${(\mathbf{M}^E)}^{-1}$ itself. To elaborate, we note that $\mathbf{L}$ is of the size $n_E^{\psi}\times n_h$, and is hence much smaller than the size $(n_h+n_E^{\psi})\times(n_h+n_E^{\psi})$ of ${(\mathbf{M}^E)}^{-1}$. Furthermore, $\mathbf{L}$, owing to the locality of the enrichment functions, is sparse. As noted earlier, $\mathbf{S}^{-1}$ is a small $n_E^{\psi}\times n_E^{\psi}$ matrix and $\mathbf{(M^{cc})}^{-1}$, being diagonal, is sparse. Since we are only interested in the action of matrix ${(\mathbf{M}^E)}^{-1}$ on a vector (as will be discussed in Section ~\ref{sec:ChFSI}), we perform the matrix-vector product using the constituent matrices without ever explicitly constructing the ${(\mathbf{M}^E)}^{-1}$ matrix. This matrix-free evaluation of any matrix-vector product presents a significant advantage for the above inversion technique over the Newton-Schultz ~\cite{Jansik2007,Nikalsson2004,Higham1997} and Taylor expansion ~\cite{Mauri1993} based methods, wherein the construction of the ${(\mathbf{M}^E)}^{-1}$ matrix is explicit and hence have huge memory requirements owing to the dense structure of  ${(\mathbf{M}^E)}^{-1}$.

Finally, we briefly compare the proposed enriched finite element method with the other existing methods which in a similar spirit seek to augment the classical finite element basis with other basis functions that efficiently capture the known physics in regions of interest. One such approach is that of partition-of-unity finite element method (PUFEM) ~\cite{Melenk1996,Babuska1997}, wherein a typical discretization can be defined as~\cite{Sukumar2009, Pask2016}
\begin{equation}
  \psi^h(\bx)=\sum_{j=1}^{n_h}N_j^C(\bx)\psi_j^C+\sum_{\alpha}^{n_E}\sum_{k=1}^{n_{PU}} N_k^{PU}(\bx)N_{\alpha}^E(\bx)\psi_{\alpha,k}^E\,,
\end{equation}
where $N_j^C(\bx)$ are the classical finite element basis functions, and $N_k^{PU}(\bx)$ is a subset of the classical finite element basis functions used to modulate the enrichment functions, $N_{\alpha}^{E,\psi}(\bx)$, thus providing a larger set of augmenting functions. Although PUFEM preserves the locality of the basis to be commensurate with conventional finite element basis, the effect of multiplying enrichment functions with a set of classical finite element basis functions results in smoother augmenting basis functions, thereby making it more prone to ill-conditioning (due to linear dependency of augmenting basis functions with classical finite element basis functions). A more serious limitation of PUFEM stems from the significant increase in the number of augmenting basis functions, which, in turn, significantly increases the size of the $\mathbf{M^{ee}}$ block of the overlap matrix $\mathbf{M}^E$, thereby making the evaluation of the $\mathbf{S^{-1}}$ in ${(\mathbf{M}^E)}^{-1}$ computationally prohibitive.

Another such approach is that of gaussian finite element mixed basis ~\cite{Yamakawa2005}, wherein a given choice of gaussian basis is used to the augment the classical finite element basis instead of atomic solutions to the Kohn-Sham problem, as used in the present work. We note that compared to the gaussian basis the atomic solutions provide a more natural choice for augmenting functions and also provide for better control over the conditioning of the basis through the use of smooth cutoff functions on the radial part of the atomic orbitals. Further, in the work on gaussian finite element mixed basis ~\cite{Yamakawa2005}, the Kohn-Sham problem was solved as a generalized eigenvalue problem using preconditioned conjugate-gradient method ~\cite{Gan2001} which is, in general, less efficient compared to the Chebyshev filtering method used in the present work, discussed subsequently.

\section{Self-consistent field iteration and Chebyshev filtering} \label{sec:ChFSI}
We begin this section with a brief outline of the well-known Kohn-Sham self-consistent field iteration (SCF) used to solve the nonlinear Kohn-Sham eigenvalue problem. This involves starting with an input guess for the charge density, $\rho_{\text{in}}$, which is used to construct the effective potential, $V_{\text{eff}}(\rho_{\text{in}},\bR)$. Having constructed $V_{\text{eff}}(\rho_{\text{in}},\bR)$, the Kohn-Sham eigenvalue problem is solved to obtain the eigenpairs ($\epsilon_i$,$\psi_i$), which are in turn used to compute the output charge density, $\rho_{\text{out}}$. If the difference between $\rho_{\text{out}}$ and $\rho_{\text{in}}$, in an appropriately chosen norm, is below a certain tolerance, then the charge density is deemed to have converged and $\rho_{\text{out}}$ denotes the ground-state charge density. Otherwise, $\rho_{\text{in}}$ is updated through a choice of mixing scheme~\cite{Anderson1965,Broyden1965,Eyert1996,Kudin2002} involving $\rho_{\text{in}}$ and $\rho_{\text{out}}$ from the current as well as those from previous iterations, and the iteration continues until convergence in charge density is achieved. 

The most computationally expensive step in every iterate of the SCF procedure is the solution of the discrete Kohn-Sham eigenvalue problem. Typically, one can use Krylov-subspace based methods such as Jacobi-Davidson~\cite{Sleijpen1996} or Krylov-Schur~\cite{Stewart2002} to evaluate the lowest few eigenpairs corresponding to the occupied eigenstates. However, benchmark studies presented in a recent work ~\cite{Motamarri2013} have shown these Krylov-subspace based methods to be about ten-fold slower in comparison to the Chebyshev filtering technique ~\cite{Zhou2006b} to compute the occupied eigenstates. Based on this relative merit, we have employed the Chebyshev filtering technique to compute the relevant eigenspectrum of the Kohn-Sham Hamiltonian. 

The key idea involved in the Chebyshev filtering approach is to progressively improve the subspace $V$ spanned by the eigenvectors of the previous SCF iteration through polynomial based power iteration to eventually compute the occupied eigenspectrum upon attaining self-consistency. It relies on two important properties of a Chebyshev polynomial $p_m(x)$ of degree $m$ to magnify the relevant (occupied) spectrum of the discrete Kohn-Sham Hamiltonian: (i) $p_m(x)$ grows rapidly outside the interval $[-1,1]$, and (ii) $|p_m(x)| \leq 1$ for $x\in[-1,1]$. For the sake of notational simplicity, we denote the discrete Kohn-Sham Hamiltonian by $\widetilde{\mathbf{H}}$, which in the classical finite element basis is ${(\mathbf{M}^C)}^{-1}\mathbf{H}^C$ and in the enriched finite element basis is ${(\mathbf{M}^E)}^{-1}\mathbf{H}^E$. The filtering technique proceeds by first mapping the unoccupied eigenspectrum of $\widetilde{\mathbf{H}}$ to $[-1,1]$ through the affine transformation $t(x) = \frac{2x-a-b}{b-a}$, where $a$ and $b$ denote the upper bounds of the occupied and unoccupied eigenspectrum of $\widetilde{\mathbf{H}}$, respectively. The upper bound of the unoccupied spectrum, $b$, is obtained inexpensively through a few Arnoldi iterations on $\widetilde{\mathbf{H}}$. The upper bound of the occupied spectrum, $a$, is obtained as the highest Rayleigh quotient of $\widetilde{\mathbf{H}}$ in the subspace $V$ of the previous SCF iteration. We denote the resultant transformed matrix as $\bar{\mathbf{H}}$. We then apply the $m$-degree Chebyshev polynomial filter $p_m(\bar{\mathbf{H}})$ on $V$ to obtain $\widetilde{V}=p_m(\bar{\mathbf{H}})V$. Owing to the rapid growth property of Chebyshev polynomials outside $[-1,1]$, the aforementioned filtering of $V$ amplifies, for each vector in $V$, the components along the eigenvectors corresponding to occupied states and damps the components along the eigenvectors corresponding to unoccupied states. The action of the Chebyshev filter on $V$ can be achieved in an efficient manner by utilizing the recursive construction of the Chebyshev polynomial ~\cite{Rivlin1990}: $p_{k+1}(x)=2xp_k(x)-p_{k-1}(x)$. Next, we orthonormalize the Chebyshev-filtered vectors to obtain the orthonormal set of vectors $Q$ spanning $\widetilde{V}$, and perform a Galerkin projection of $\widetilde{\mathbf{H}}$ onto $\widetilde{V}$ to obtain the following reduced generalized eigenvalue problem,
\begin{equation}
Q^T\mathbf{H}Q\Psi_i=\epsilon_iQ^T\mathbf{M}Q\Psi_i\,,
\end{equation}
where $\{\mathbf{H}$, $\mathbf{M}$, $\epsilon_i\}$ represent $\{\mathbf{H}^C$, $\mathbf{M}^C$, $\epsilon_i^C\}$ or $\{\mathbf{H}^E$, $\mathbf{M}^E$, $\epsilon_i^E\}$ corresponding to the classical or enriched finite element discretization, respectively. We can now solve the above generalized eigenvalue problem, whose dimension is commensurate with the number of electrons in the system, using direct solvers to obtain the eigenvalues $\epsilon_i$ and their corresponding projected eigenvectors $\Psi_i$. We subsequently rotate the projected eigenvectors to the original space to obtain the eigenvectors $Q\Psi_i$, which along with the eigenvalues $\epsilon_i$ are used to evaluate the charge density. Finally, the subspace $V$ is updated to $\widetilde{V}$ for the next SCF iteration. 

We remark that in order to gain computational efficiency, we exploit the elemental structure in $\mathbf{H}^E$ (or $\mathbf{H}^C$) and ${(\mathbf{M}^E)}^{-1}$ (or ${(\mathbf{M}^C)}^{-1}$) to perform the matrix-vector products involved in the evaluation of $\widetilde{V}=p_m(\bar{\mathbf{H}})V$. To elaborate, we consider the case of enriched finite element and note that all the blocks in the $2\times2$ block structure of $\mathbf{H}^E$ and all the constituent matrices (except ${(\mathbf{M^{cc}})}^{-1}$ and $\mathbf{S}^{-1}$) can be constructed, owing to the locality of the basis, by assembling contributions from individual elements. However, since we are interested only in the action of these matrices on vectors, we perform the matrix-vector products by first evaluating elemental matrix-vector products and then assembling the resultant elemental vector, without explicitly assembling any global matrix.  We also note that, the dimension of the subspace $V$, denoted by $N$, is chosen to be greater than the number of occupied orbitals so as to avoid numerical instabilities for systems with small band-gaps or degenerate energy levels close to the Fermi energy, and also to avoid missing out any occupied eigenstate between two successive SCF iterations. Typically, we choose $N \sim \frac{N_e}{2}+20$. Further, we note that Kohn-Sham orbitals of single atoms represent a good initial guess for the subspace $V$ for the first SCF iteration, and is adopted in the present work. 

We note that the degree $m$ of the Chebyshev polynomial filter needed to obtain a good approximation to the occupied eigenspace of the Kohn-Sham Hamiltonian depends on: (i) the separation between eigenvalues in the occupied part of the eigenspectrum, and (ii) the ratio between the spectral widths of the occupied and unoccupied part of the eigenspectrum of $\widetilde{\mathbf{H}}$. While the separation between the occupied eigenvalues depends on the materials system, the ratio of the spectral widths of the occupied and unoccupied parts of the eigespectrum depends on the largest eigenvalue of $\widetilde{\mathbf{H}}$, which, in turn depends on the finite-element discretization---it increases with decreasing element size. Typically, in a pseudopotential calculation, where the orbitals and the electrostatic potentials vary smoothly, one can use a relatively coarse finite element mesh to achieve chemical accuracies of $\sim1\,$meV per atom using the classical finite element method. For such coarse finite element discretizations, a Chebyshev polynomial degree between 10 to 50 is sufficient to compute the occupied eigenspace. However, in all-electron calculations, where the orbitals are characterized by sharp variations near atomic cores and the external potential has Coulomb-singularity, one requires a highly refined finite element mesh near the atomic cores to achieve chemical accuracies of $\sim1\,$mHa per atom. In addition to the significant increase in the degrees of freedom, such mesh refinement also increases the upper bound of the unoccupied eigenspectrum, thereby requiring a very high Chebyshev polynomial degree, $\order(10^3)$, to effectively compute the occupied eigenspace. These shortcomings of the classical finite element discretization in the context of all-electron calculations are noted in ~\cite{Motamarri2013}, where comparisons were made against plane-wave basis for pseudopotential calculations and against gaussian basis for all-electron. It was noted that while the classical finite elements basis outperforms the plane-wave basis in pseudopotential calculations on the benchmark systems studied, they were ten-fold slower in comparison to the gaussian basis in all-electron calculations. These disadvantages of the classical finite element basis for all-electron calculations are mitigated by using the proposed enriched finite element basis, as will be demonstrated in the subsequent section.       

\section{Results and Discussion} \label{sec:Results}
In this section, we discuss the rate of convergence, accuracy, performance, and parallel scalability of the proposed enriched finite method for all-electron calculations. We first study the rate of convergence of ground-state energy with respect to element size for methane and carbon monoxide molecules. We then demonstrate the accuracy and performance of the enriched finite element method using large-scale non-periodic semi-conducting and heavy metallic systems. We use non-periodic silicon nano-clusters of various sizes, with the largest one containing 621 atoms (8694 electrons), as our benchmark semi-conducting systems. For heavy metallic systems, we use gold nano-clusters, $\text{Au}_n (n=1-23)$, as our benchmark systems. In order to assess the accuracy, reduction in degrees of freedom, reduction in Chebyshev polynomial degree, and performance of the enriched finite element method, we use, wherever possible, the classical finite element method as a reference. Depending upon the system, we use spectral hexahedral finite elements of polynomial order 2 to 6, denoted as HEX27, HEX64SPEC, HEX125SPEC, HEX216SPEC, HEX343SPEC respectively. We also compare, wherever possible, the accuracy and performance of the enriched finite element method with that of gaussian basis. All our calculations using the gaussian basis are performed with the NWCHEM ~\cite{Nwchem2010} package. We use n-stage Anderson mixing ~\cite{Anderson1965} for charge density mixing in all our enriched and classical finite element method based calculations. Finally, we present the parallel scalability of our implementation of the proposed enriched finite element method using Message Passing Interface (MPI). The scalability studies as well as the benchmark studies demonstrating the computational efficiency, reported subsequently, are conducted on a parallel computing cluster with the following configuration: Intel Xeon E5-2680v3 CPU nodes with 24 processors (cores) per node, 128 GB memory per node, and Infiniband networking between all nodes for fast MPI communications.  
\subsection{Rate of convergence}
In this section, we study the rate of convergence of the ground-state energy with element size, $h$, using quadratic (HEX27) and cubic (HEX64SPEC) spectral finite elements. To this end, we generate a sequence of finite element meshes with increasingly smaller element sizes by uniformly subdividing the coarsest mesh. The ground-state energy, $E_h$, obtained from each of the HEX64SPEC meshes are used in the expression
\begin{equation}
  |E_h-E_0|=Ch^{q}
\end{equation}
to compute the constants $E_0$, $q$ and $C$ through a least-square fit. In the above expression, $E_0$ is the extrapolated continuum ground-state energy obtained as $h \rightarrow 0$. We use the $E_0$ obtained from HEX64SPEC to compute the relative error $\frac{|E_h-E_0|}{|E_0|}$ for both HEX27 and HEX64SPEC meshes. To assess the accuracy of $E_0$, we also compare it against the ground-state energy obtained using the polarization consistent-4 (pc-4) ~\cite{Jensen2002} gaussian basis.

\begin{figure} [htbp] 
  \centering
  \includegraphics[width=1.0\columnwidth]{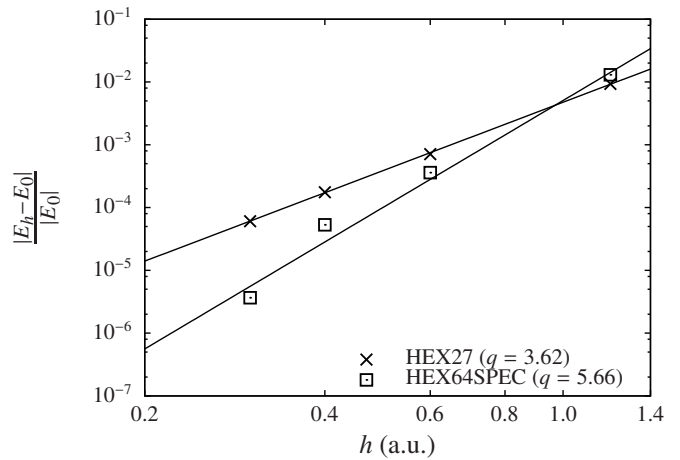}
  \caption{Convergence of energy with respect to element size for methane molecule}
  \label{fig:CH4}
\end{figure}
\begin{figure} [htbp] 
  \centering
  \includegraphics[width=1.0\columnwidth]{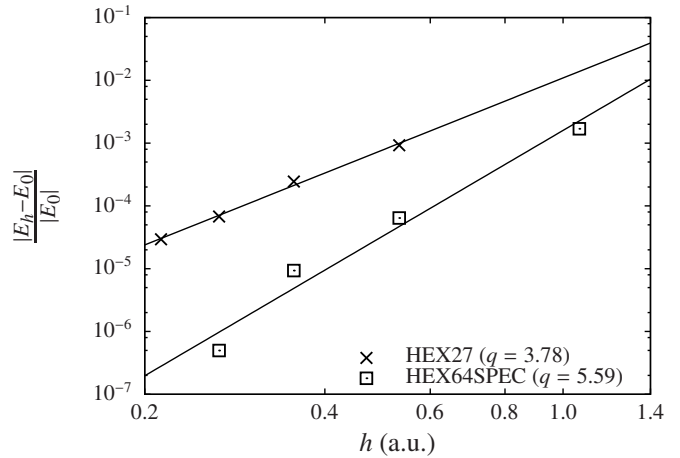}
  \caption{Convergence of energy with respect element size for carbon monoxide molecule}
  \label{fig:CO}
\end{figure}

For the benchmark systems in our convergence study, we consider two systems: (i) methane molecule with a C-H bond length of $2.0784\,$a.u. and H-C-H bond angle of 109.4712$^{\circ}$, and (ii) carbon monoxide molecule with a C-O bond length of $2.1297\,$a.u.. For both the systems, we use a Chebyshev filter of order 60 to compute the occupied eigenspace and Fermi-Dirac smearing with $T=500K$. For methane, the value of $E_0$ is evaluated to be $-40.11993$~Ha and the ground-state energy from pc-4 gaussian basis is $-40.11992$~Ha. For carbon monoxide, the value of $E_0$ is evaluated to be $-112.47189$~Ha and the ground-state energy from pc-4 gaussian basis is $-112.47188$~Ha. Next, we plot the relative error, $\frac{|E_h-E_0|}{|E_0|}$, against the smallest element size, and obtain the rates of convergence from the slopes of these plots. As evident from Figures ~\ref{fig:CH4} and ~\ref{fig:CO}, we obtain close to optimal rates of convergence in energy of $\mathcal{O}(h^{2k})$, where $k$ is the polynomial order ($k=2$ for HEX27 and $k=3$ for HEX64SPEC). The results also suggest higher accuracies obtained with HEX64SPEC when compared to HEX27 for the same mesh size. We note that the numerically obtained rates of convergence deviate slightly from the theoretically optimal rates due to other numerical errors---beyond the discretization errors in the theoretical estimates---that are present in simulations, such as quadrature errors, errors due to stopping tolerance in the iterative solutions of the Poisson problem, diagonalization of the Hamiltonian and the self consistent field iteration.

\subsection{Large-scale materials systems}
We now discuss the accuracy and performance of the proposed enriched finite element method using large-scale semi-conducting and heavy metallic materials systems. We also compare, wherever possible, the proposed method against classical finite element and gaussian basis based calculations. 
\subsubsection{Semi-conducting systems: Silicon nano-clusters}
We consider silicon nano-clusters of various sizes, containing $1\times1\times1 \;(252\;\text{electrons})$, $2\times1\times1 \;(434\;\text{electrons})$, $2\times2\times2 \;(1330\;\text{electrons})$, $3\times3\times3\;(3920\;\text{electrons})$; and  $4\times4\times4\;(8694\;\text{electrons})$ diamond unit cells, as our benchmark semi-conducting systems. We employ a lattice constant of $10.26\,$a.u. in our calculations. These are isolated clusters in vacuum and we do not use any surface passivation. To obtain the characteristic element size to be used in the enriched finite element based calculations of the nano-clusters, we first obtain the reference ground-state energy, $E_{ref}$, for a single silicon atom by solving its 1D-radial Kohn-Sham eigenvalue problem as mentioned in Section ~\ref{sec:EFEMDiscretize}. Next, we choose a fourth-order (HEX125SPEC) finite element mesh for which the single atom ground-state energy obtained from the enriched finite element based calculation is within $1\,$mHa accuracy with respect to $E_{ref}$. Similarly, to obtain the characteristic element size for the classical finite element based calculations of the nano-clusters, we choose a fifth-order (HEX216SPEC) finite element mesh which is also within a $1\,$mHa accuracy for the single atom ground-state energy. We note that the smallest element size, thus obtained for the classical finite element based calculation is found to be an order of magnitude smaller than that of the smallest element size obtained in the enriched finite element based calculation. This, in turn, affects the largest eigenvalue of the Kohn-Sham Hamiltonian which is found to be $\order(10^6)$ in case of classical finite elements, thereby, requiring a Chebyshev polynomial filter of degree $1500$ to compute the occupied eigenstates. Correspondingly, for the enriched finite element case, the largest eigenvalue is found to be $\order(10^3)$, thereby, requiring a $\sim20$-fold smaller Chebyshev polynomial degree of $80$ to compute the occupied eigenstates. These choices for element sizes and Chebyshev polynomial degrees from single atom calculations are then carried forward to the nano-cluster calculations. We note that owing to the steep computational demand arising from large number of basis functions and high Chebyshev polynomial degree in the case of classical finite element based all-electron calculations, we could only perform studies up to $2\times2\times2$ nano-cluster size using the computational resources available to us. We also compare the accuracy and performance of the enriched finite element method with gaussian basis. We use the polarization consistent (pc) family of gaussian basis as it provides a hierarchy of increasingly larger basis sets. Specifically, we use pc-3 and pc-4 basis as they are both commensurate with the $\sim1\,$mHa accuracy when compared with aforementioned $E_{ref}$ for a single silicon atom. All the calculations with gaussian basis are performed using Direct Inversion of Iterative Subspace (DIIS) ~\cite{Pulay1982} as well as the quadratically convergent minimization scheme ~\cite{Bacskay1981}, both available within the NWCHEM package, and the computational time from the more efficient scheme is reported. For the DIIS scheme, extrapolation of up to 10 Fock matrices were used. Table ~\ref{tab:SiCFEEFE} compares the ground-state energy, degrees of freedom (number of basis functions) per atom and the total computation CPU time (CPU time = number of cores $\times$ wall-clock time) for various cluster sizes using classical and enriched finite element basis. Similarly, Table ~\ref{tab:SiEFEPC} compares the ground-state energy and the total computation CPU time for various cluster sizes using enriched finite element, pc-3 and pc-4 basis. In all these calculations, we used a Fermi-Dirac smearing with $T=500K$. 

\begin{table}[h!]
  \caption{Comparison of classical and enriched finite element (FE) basis: Energy per atom ($E$ in Ha), degrees of freedom per atom (DoF), and total computation CPU time (in CPU hours) for various silicon nano-clusters.} 
\begin{tabular}{M{0.25\columnwidth}M{0.35\columnwidth}M{.35\columnwidth}}
\hline 
\hline
  Si $1\times1\times1$ & Classical FE & Enriched FE\\ \hline
  $E$ & $-288.320035$ & $-288.319450$ \\
  DoF & $402,112$ & $14,728$ \\
  CPU Hrs & $1599.15$ & $24.81$ \\ \hline

  Si $2\times1\times1$ & Classical FE & Enriched FE\\ \hline
  $E$ & $-288.334123$ & $-288.333872$ \\
  DoF & $386,205$ & $13,557$\\ 
  CPU Hrs & $16441.43$ & $57.10$ \\ \hline

  Si $2\times2\times2$ & Classical FE & Enriched FE\\ \hline
  $E$ & $-288.359459$ & $-288.359266$ \\
  DoF & $360,467$ & $10,642$\\ 
  CPU Hrs & $75936.4$ & $553.13$ \\ \hline \hline
\end{tabular}
  \label{tab:SiCFEEFE}
\end{table}
\begin{table}[h!]
  \caption{Comparison of enriched finite element, pc-3 and pc-4 basis: Energy per atom ($E$ in Ha) and total computation CPU time (in CPU hours) for various silicon nano-clusters.} 
  \begin{tabular}{M{0.2\columnwidth}M{0.26\columnwidth}M{.26\columnwidth}M{0.26\columnwidth}}
\hline 
\hline
  Si $1\times1\times1$ & Enriched FE & pc-3 & pc-4\\ \hline
  $E$ & $-288.319450$ & $-288.318996$ & $-288.319448$ \\
    CPU Hrs & $24.81$ & $8.39$ & $98.88$ \\ \hline

  Si $2\times1\times1$ & Enriched FE & pc-3 & pc-4\\ \hline
  $E$ & $-288.333872$ & $-288.333447$ & $-288.333898$\\
  CPU Hrs & $57.10$ & $151.74$ & $1817.30$\\ \hline

  Si $2\times2\times2$ & Enriched FE & pc-3 & pc-4\\ \hline
    $E$ & $-288.359266$ & $-288.360045$ & FTC\footnote{FTC: Failed to converge}\\
  CPU Hrs & $553.13$ & $4097.29$ & $-$\\ \hline \hline

  Si $3\times3\times3$ & Enriched FE & pc-3 & pc-4\\ \hline
  $E$ & $-288.374721$ & FTC & FTC\\
  CPU Hrs & $6252.15$ & $-$ & $-$\\ \hline

  Si $4\times4\times4$ & Enriched FE & pc-3 & pc-4\\ \hline
  $E$ & $-288.381425$ & FTC & FTC\\
  CPU Hrs & $45053.82$ & $-$ & $-$\\ \hline \hline
\end{tabular}
  \label{tab:SiEFEPC}
\end{table}

As is evident from Tables ~\ref{tab:SiCFEEFE} and ~\ref{tab:SiEFEPC}, the enriched finite element basis achieves accuracies of within $1\,$mHa in the ground-state energies per atom when compared with classical finite element, pc-3 and pc-4 basis. We observe a staggering $60-$ to $300-$fold reduction in the total computation CPU time for the enriched finite element basis when compared with the classical finite element basis. This reduction in computation time stems from a $\sim 30-$fold reduction in the degrees of freedom as well as a $\sim 20-$fold reduction in the Chebyshev polynomial degree as compared to the classical finite element basis. When compared with the pc-3 gaussian basis, the enriched finite element is a factor $\sim3$ slower in the case of the smallest ($1\times1\times1$) cluster. However, it outperforms the pc-3 basis, in total computation CPU time, by a factor of $2.5$ for the $2\times1\times1$ cluster and by a factor of $7.5$ for the $2\times2\times2$ cluster. Similarly, the enriched finite element basis outperforms the pc-4 gaussian basis by factors $4$ and $30$ for the $1\times1\times1$ and $2\times1\times1$ clusters, respectively. We note that the pc-3 basis failed to converge for the $3\times3\times3$ and $4\times4\times4$ clusters, whereas the pc-4 basis failed to converge for $2\times2\times2$ and higher clusters. The failure of the pc-3 and pc-4 basis to converge for larger system sizes is primarily due to linear dependency of the gaussian basis functions for larger system sizes. These results suggest that the enriched finite element basis offers a computationally efficient and robust basis for all-electron calculations in semi-conducting systems as compared to both classical finite element and gaussian basis.     

\subsubsection{Heavy metallic systems: Gold nano-clusters}
Next, we consider gold nano-clusters, $\text{Au}_n (n=1-23)$, to study the accuracy and performance of the enriched finite element basis. For $n=2$ and $n=6$, we use the stable geometries obtained in a previous work ~\cite{Wang2002} wherein the $\text{Au}_2$ has a bond length of $4.818\,$a.u. and $\text{Au}_6$ has a planar triangle geometry with $D_{3h}$ symmetry and bond length of $5.055\,$a.u.. The $\text{Au}_{14}$ and $\text{Au}_{23}$ nano-clusters were constructed as $1\times1\times1$ and $2\times1\times1$ face centered cubic (FCC) lattice, respectively, with a lattice constant of $6.812\,$a.u.. We follow the same strategy as used for silicon nano-clusters to obtain the characteristic element sizes and Chebyshev polynomial degrees that are to be used in gold nano-cluster calculations, both using classical and enriched finite element basis. We use fifth-order (HEX216SPEC) and sixth-order (HEX343SPEC) finite elements for the enriched and classical finite element based calculations, respectively. We note that since gold is much heavier than silicon, it is characterized by more sharply oscillating orbitals and much steeper electrostatic potentials in comparison to silicon, thereby requiring smaller element sizes than those used in silicon to achieve similar accuracy. This, in turn, results in an increment in the largest eigenvalues of the Hamiltonian, which are found to be $\order(10^4)$ and $\order(10^8)$, for the enriched and classical finite element basis, respectively, thereby requiring higher Chebyshev polynomial degrees to accurately compute the occupied eigenstates. We note that the Chebyshev polynomial based filtering technique, being analogous to the power iteration method, can generate an ill-conditioned space for a very high polynomial degree, thereby resulting in numerical issues. To circumvent this, we employ, at each SCF iteration, multiple passes of a low polynomial degree Chebyshev filter and orthonormalize the filtered vectors between two successive passes. For all our gold cluster calculations based on the enriched finite element basis we used $30$ passes of a Chebyshev filter of degree $20$, whereas $10$ passes of a Chebyshev filter of degree $1200$ have been employed for the classical finite element based calculations. We note that in the case of classical finite element based calculations, owing to the huge computational cost, we could perform calculations only up to $\text{Au}_2$ using the computational resources at our disposal. Further, we do not present a comparison with gaussian basis owing to the lack of a good hierarchical non-relativistic basis for gold. Table ~\ref{tab:AuCFEEFE} presents the comparison of the ground-state energies, degrees of freedom and total computation CPU times for the gold nano-clusters using classical and enriched finite element basis.

\begin{table}[!]
  \caption{Comparison of classical and enriched finite element (FE) basis: Energy per atom ($E$ in Ha), degrees of freedom per atom (DoF), and total computation CPU time (in CPU hours) for various gold nano-clusters.} 
\begin{tabular}{M{0.25\columnwidth}M{0.35\columnwidth}M{.35\columnwidth}}
\hline 
\hline
  $\text{Au}_1$ & Classical FE & Enriched FE\\ \hline
  $E$ & $-17860.7623$ & $-17860.7622$ \\
  DoF & $5,040,409$ & $120,361$ \\
  CPU Hrs & $612.22$ & $43.39$ \\ \hline

  $\text{Au}_2$ & Classical FE & Enriched FE\\ \hline
  $E$ & $-17860.8001$ & $-17860.8019$ \\
  DoF & $4,659,399$ & $122,300$\\ 
  CPU Hrs & $22950.25$ & $220$ \\ \hline

  $\text{Au}_6$ & Classical FE & Enriched FE\\ \hline
  $E$ & $-$ & $-17860.8249$ \\
  DoF & $-$ & $178,906$\\ 
  CPU Hrs & $-$ & $1924.42$ \\ \hline

  $\text{Au}_{14}$ & Classical FE & Enriched FE\\ \hline
  $E$ & $-$ & $-17860.8077$ \\
  DoF & $-$ & $88,657$\\ 
  CPU Hrs & $-$ & $3740.29$ \\ \hline
 
 $\text{Au}_{23}$ & Classical FE & Enriched FE\\ \hline
  $E$ & $-$ & $-17860.8045$ \\
  DoF & $-$ & $80,397$\\ 
  CPU Hrs & $-$ & $8171.40$ \\ \hline
\end{tabular}
  \label{tab:AuCFEEFE}
\end{table}

As is evident from Table ~\ref{tab:AuCFEEFE}, the enriched finite element basis obtains chemical accuracy in the ground-state energies per atom with far fewer degrees of freedom. In terms of computational efficiency, while the enriched finite element basis achieves $\sim14-$fold speedup over the classical finite $\text{Au}_1$, we observe $\sim100-$fold speedup for $\text{Au}_2$. Once again, these speedups for the enriched finite element basis are the result of a $40-$fold reduction in the number of degrees of freedom and a $20-$fold reduction in the Chebyshev polynomial degree as compared to that of the classical finite element basis. These numerical experiments demonstrate the accuracy and efficiency for all-electron calculations in heavy metallic systems.
\subsection{Scalability}
\begin{figure}[h!] 
  \centering
  \includegraphics[width=1.0\columnwidth]{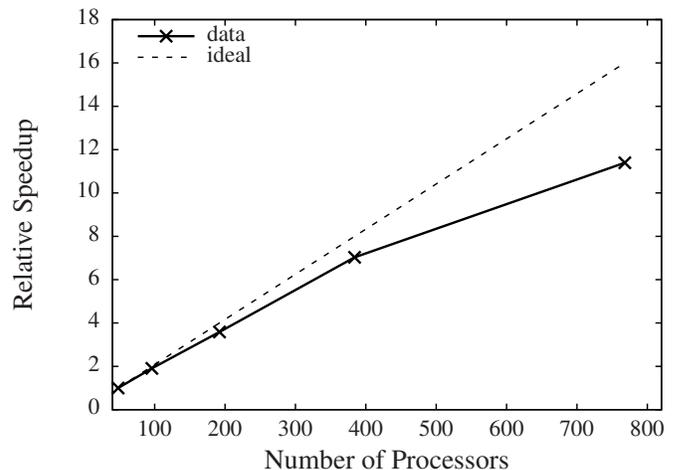}
  \caption{Parallel scalability of the enriched finite element implementation.}
  \label{fig:Scalability}
\end{figure}

We demonstrate the parallel scalability (strong scaling) of the proposed enriched finite element basis in Figure ~\ref{fig:Scalability}. We choose the $3\times3\times3$ silicon nano-cluster containing $\sim4$ million degrees of freedom (number of basis functions) as our fixed benchmark system and report the relative speedup with respect to the wall time on 48 processors. The use of any number of processors below 48 was infeasible owing to the memory requirement posed by the system. As evident from the figure, the scaling is in good agreement with the ideal linear scaling behavior up to 384 processors, at which we observe a parallel efficiency of 87.8$\%$. However, we observe a considerable deviation from linear scaling behavior at 768 processors with a parallel efficiency of 71.2$\%$. This is attributed to the fact that at 768 processors the number of degrees of freedom possessed by each processor falls below 5000, which is too low to achieve good parallel scalability.  

\section{Summary} \label{sec:Summary}
We have developed a computationally efficient mixed basis, termed as enriched finite element basis, for all-electron DFT calculations which combines the efficiency of atomic-orbitals-type basis to capture the sharp variations of the electronic fields closer to the atoms and the completeness of the classical finite element basis. This work demonstrates the marked computational advantage afforded by the enriched finite element basis over the classical finite element basis for all-electron DFT calculations.

The proposed method is developed based on the following key ideas. Firstly, we augmented the classical spectral finite element basis with enrichment functions constructed from single-atom Kohn-Sham orbitals and electrostatic potentials. The enrichment functions are inexpensively pre-computed and stored by solving radial Kohn-Sham equations for all atoms in the periodic table. The enrichment functions are instrumental in capturing the sharp variations of the Kohn-Sham orbitals close to an atom, thereby, mitigating the need of high mesh refinement near the atomic cores. Secondly, we used smooth cutoff functions to truncate the enrichment functions so as to ensure locality as well as better conditioning of the enriched finite element basis. Thirdly, we employ a divide and conquer strategy to construct an adaptive quadrature grid to efficiently evaluate the integrals involving the enrichment functions. Next, in order to convert the generalized Kohn-Sham eigenvalue problem to a standard eigenvalue problem, we employed a computationally efficient scheme to evaluate the inverse of the overlap matrix in the enriched finite element basis, by exploiting the block-wise matrix inversion. The use of spectral finite elements along with Gauss-Lobatto-Legendre quadrature rule is crucial in rendering the classical-classical block of the overlap matrix diagonal, whereas the use of the block-wise matrix inversion is crucial in utilizing the sparsity of the constituent matrices in the inverse of the overlap matrix for an efficient evaluation of the ensuing matrix-vector products. Finally, we employed the Chebyshev polynomial based filter to compute the occupied eigenstates. Here, we exploited the finite element structure in the Hamiltonian and the inverse overlap matrices to achieve an efficient and scalable implementation of the matrix-vector products involved in the action of the Chebyshev filter on a subspace.

In terms of the numerical convergence afforded by the enriched finite element basis, we demonstrated close to optimal rates of convergence for the ground-state energy with respect to the finite element discretization. We demonstrated the accuracy and performance of the proposed enriched finite element basis on: (i) silicon nano-clusters of various sizes, with the largest cluster containing 8694 electrons; and (ii) gold nano-clusters of various sizes, with the largest cluster containing 1817 electrons. We obtained good agreement in the ground-state energies when compared to classical finite element and gaussian basis. In the larger clusters considered in this study, the enriched finite element basis provides a staggering $50-300$ fold speedup compared to the classical finite element basis, which is attributed to a $30-$fold reduction in the degrees of freedom as well as a $20-$fold reduction in the Chebyshev polynomial degree. We also observed a significant outperformance by the enriched finite element basis relative to gaussian basis (pc-3 and pc-4). Furthermore, we were able to perform ground-state energy calculations for silicon clusters containing 280 and 621 atoms, for which the gaussian basis failed to converge owing to linear dependency of the basis. In terms of parallel scalability, we obtained good parallel efficiency with almost linear scaling up to 384 processors for the benchmark system comprising of 280 atoms silicon nano-cluster and containing $\sim 4$ million basis functions.

The proposed method offers a computationally efficient, systematically improvable, and scalable basis for large scale all-electron DFT calculations, applicable to both light and heavy atoms. The use of the enrichment in developing linear-scaling DFT algorithms for all-electron calculations based on finite element basis ~\cite{Motamarri2014,Motamarri2016b} or Tucker-tensor basis~\cite{Motamarri2016} holds good promise, and is currently being investigated. Furthermore, the use of enrichment ideas in conjunction with reduced-order scaling DFT algorithms can also be effectively utilized in the evaluation of the exact exchange operator, and forms a future direction of interest. Last, but not the least, the enriched finite element basis can be useful in a systematic study of the applicability and accuracy of various pseudopotential approximations on a wide range of materials and external conditions.
\acknowledgements
{We thank Dr. Motamarri for useful discussion during the course of this work. We gratefully acknowledge the support from Army Research Office through Grant No. W911NF-15-1-0158, and the support from National Science Foundation through Grant No. 1053145, under the auspices of which part of this work was conducted. This work used the Extreme Science and Engineering Discovery Environment (XSEDE), which is supported by National Science Foundation grant number ACI-1053575. We also acknowledge Advanced Research Computing at University of Michigan for providing computing resources through the Flux computing platform.
}
\bibliographystyle{apsrev4-1}
\end{document}